\documentclass[prl,reprint,nofootinbib,showpacs]{revtex4-1}
\usepackage{amsmath, amssymb,bm,xcolor,graphicx,mathtools,enumitem}
\usepackage[colorlinks=true,citecolor=blue]{hyperref}
\usepackage[T1]{fontenc}

\newcommand\redsout{\bgroup\markoverwith{\textcolor{red}{\rule[0.5ex]{2pt}{0.4pt}}}\ULon}

\def\be{\begin{equation}}
\def\ee{\end{equation}}

\begin{document}
\title{Self-forces in arbitrary dimensions}

\author{Abraham I. Harte$^1$, Peter Taylor$^{1}$ and \'Eanna \'E. Flanagan$^{2,3}$}
\affiliation{$^1$Centre for Astrophysics and Relativity, School of Mathematical Sciences\\ Dublin City University, 
Glasnevin, Dublin 9, Ireland}
\affiliation{$^2$Cornell Center for Astrophysics and Planetary Science, Cornell
University, Ithaca, NY 14853}
\affiliation{$^3$Department of Physics, Cornell University, Ithaca, NY 14853}

%
%
\newcount\hh
\newcount\mm
\mm=\time
\hh=\time
\divide\hh by 60
\divide\mm by 60
\multiply\mm by 60
\mm=-\mm
\advance\mm by \time
\def\hhmm{\number\hh:\ifnum\mm<10{}0\fi\number\mm}

\begin{abstract}
Bodies coupled to electromagnetic or other long-range fields are subject to radiation reaction and other effects in which their own fields can influence their motion. Self-force phenomena such as these have been poorly understood for spacetime dimensions not equal to four, despite the relevance of differing dimensionalities for holographic duals, effectively two-dimensional condensed matter and fluid systems, and so on. We remedy this by showing that forces and torques acting on extended electromagnetic charges in all dimensions $d \geq 3$ have the same functional form as the usual test body expressions, except that the electromagnetic field appearing in those expressions is not the physical one; it is an effective surrogate. For arbitrary even $d \geq 4$, our surrogate field locally satisfies the source-free field equations, and is conceptually very similar to what arises in the Detweiler-Whiting prescription previously established when $d=4$. The odd-dimensional case is different, involving effective fields which are not necessarily source-free. Moreover, we find a 1-parameter family of natural effective fields for each odd $d$, where the free parameter---a lengthscale---is degenerate with (finite) renormalizations of a body's stress-energy tensor. While different parameter choices can result in different forces, they do so without affecting physical observables. Having established these general results, explicit point-particle self-forces are derived in odd-dimensional Minkowski spacetimes. Simple examples are discussed for $d=3$ and $d=5$, one of which illustrates that the particularly slow decay of fields in three spacetime dimensions results in particles creating their own ``preferred rest frames:'' Initially-static charges which are later perturbed have a strong tendency to return to rest. While the focus here is on the electromagnetic self-force problem, all results are easily carried over to the scalar and gravitational cases.

\end{abstract}
\vskip 1pc

\maketitle

The motion of small bodies is central to some of the most enduring problems in physics. If such a body is coupled to an electromagnetic, gravitational, or other long-range field, it may be subject to net forces exerted by its own contributions to that field. This ``self-force'' strongly influences, for example, charged particles circulating in particle accelerators and the shrinking orbits of black hole binaries due to the emission of gravitational radiation. Despite its initial appearance of simplicity, the study of self-interaction involves a number of physical and mathematical subtleties. This has led to more than a century of literature on the electromagnetic self-force; see \cite{Abraham, Lorentz, Dirac, DeWittBrehme, Nodvik, Spohn, Yaghjian, GrallaHarteWald, HarteEM}.

While impetuses for these works have varied considerably, the past two decades have seen a concerted effort---motivated largely by gravitational wave astronomy---to understand the gravitational self-force problem in general relativity. This has led to a number of theoretical and computational advances which considerably improve our understanding of self-interaction, in both the gravitational and electromagnetic contexts \cite{PoissonLR, HarteReview, PoundReview, WardellReview}. Separately, new aspects of the electromagnetic self-force are also beginning to be accessible to investigation via high-power laser experiments \cite{laserRev, Laser2}.

Here, we move beyond the existing literature to rigorously study self-interaction in different numbers of dimensions. There are three reasons for this: First, considerations in different numbers of dimensions refines our understanding of precisely what is important and what is not; lessons learned in this way may significantly inform future considerations even in four dimensions, particularly in more complicated theories which have not yet been understood. Second, considerations of theories in non-physical numbers of dimensions can, via holographic dualities, be related to ordinary four-dimensional systems; for example, the five-dimensional self-force might be used to understand jet quenching in four-dimensional dimensional quark-gluon plasmas \cite{Shuryak2}. 

Our final reason for considering different numbers of dimensions is that the self-force in three spacetime dimensions may be proportionally stronger than in four dimensions, both in terms of instantaneous magnitude \cite{HFTPaper1} and---as argued below---in the particularly slow decay of fields which encode a kind of ``memory'' of a system's past. Moreover, systems in which is this is relevant may be accessible to experiment. For example, ``pilot wave hydrodynamics'' involves a number of striking phenomena observed to be associated with oil droplets bouncing on a vibrating bath \cite{PilotWave}. Each bounce generates surface waves on the bath, but these waves also affect the horizontal motion of the droplet. This type of feedback with a long-range field (the surface waves) is reminiscent of a self-force problem in two spatial dimensions. Separately, there are a variety of condensed matter systems which act as though they are confined to one or two spatial dimensions \cite{Superfluid, TopIns}.

We do not consider any particular fluid or condensed matter system in this Letter, but instead explore a standard electromagnetic self-force problem in different numbers of dimensions. To the best of our knowledge, the literature does not contain any rigorous derivations of self-force other than in four dimensions, except for recent work restricted to static bodies \cite{HFTPaper1}. The dynamical case considered here is considerably different and more rich. Moreover, we find qualitative differences between charged-particle behavior in even and odd numbers of dimensions.

Our strategy is not to obtain a ``point particle self-force'' as any kind of fundamental object, but instead to derive laws of motion for extended objects, and then to evaluate point-particle limits of those laws. Although we focus for concreteness on the electromagnetic self-force problem, analogous results are easily obtained for the scalar and (at least first order) gravitational self-force. In the electromagnetic context, a compact body in a $d$-dimensional spacetime $(M,g_{ab})$ is associated with a conserved current density $J_a$, and the electromagnetic field $F_{ab}$ in a neighborhood of that body satisfies the Maxwell equations
\begin{equation}
	\nabla_{[a} F_{bc]} = 0, \qquad \nabla^b F_{ab} = \omega_{d-1} J_a,
\end{equation}
where $\omega_{D} \equiv 2 \pi^{D/2} / \Gamma(D/2)$ is chosen by convention to be the area of a unit sphere in $\mathbb{R}^D$. We also assume that the system's stress-energy tensor is conserved, $\nabla_b T^{ab} = 0$, and that it reduces immediately outside the body of interest to the standard expression $T^{ab}_\mathrm{em}[F_{cd}]$ for the stress-energy associated with $F_{ab}$ in vacuum. Using that expression also in the interior of the body, the body's ``own'' stress-energy, $T^{ab}_\mathrm{body} \equiv T^{ab} - T^{ab}_\mathrm{em}$, must satisfy $\nabla_b T^{ab}_\mathrm{body} = F_{ab} J^b$. Every portion of an extended charge is thus acted upon by the Lorentz force density $F_{ab} J^b$.

The question we now ask is how this density ``integrates up'' to affect a body's overall motion. One difficulty is that $F_{ab} J^b$ depends nonlinearly and nonlocally on $J_a$, and can be almost arbitrarily complicated. Despite this, experience suggests that there are physically-interesting regimes in which the (appropriately-defined) \textit{net} force is not complicated at all, implying laws of motion which are effectively universal. Deriving this universality and making it precise is the essence of the self-force problem. 

The approach adopted here uses a formalism developed by one of us \cite{HarteScalar, HarteEM, HarteGrav, HarteReview}, which provides a rigorous framework with which to analyze problems of motion in a wide variety of contexts. Crucially, most of that framework is agnostic to the number of dimensions. To review it briefly, one first defines ``bare'' linear and angular momenta for an extended body as particular integrals over the body's interior; we take these definitions to coincide with Dixon's \cite{Dix70b, Dix74, HarteReview}. Next, stress-energy conservation is used to derive forces and torques---rates of change of the momenta---which result in integrals for each force or torque component. Rather than displaying these integrals explicitly, we note that each force or torque component is a bilinear functional with the form $\mathcal{F}_\tau[ F_{ab}, J_c ]$, where $\tau$ provides a parametrized notion of time. Unless self-fields are negligible, these functionals are difficult to analyze as-is. The formalism we use nevertheless provides a set of tools which allows one to easily derive that
\begin{equation}
	\mathcal{F}_\tau [ F_{ab} , J_c ] = \mathcal{F}_\tau [ \hat{F}_{ab}, J_c ] - \delta \mathcal{T}_\tau[ J_a ]
	\label{fVary}
\end{equation}
for a wide class of nonlocal transformations $F_{ab} \mapsto \hat{F}_{ab}$. The difference term $\delta \mathcal{T}_\tau [ J_a ]$ depends nonlinearly on its argument and has properties which allow it to be absorbed into a redefinition, or finite renormalization, of a body's stress-energy tensor. In flat spacetime, these renormalizations are relevant only for the body's linear and angular momenta. More generally, quadrupole and higher higher moments of the renormalized stress-energy tensor can also arise in the laws of motion, where they appear in gravitational forces and torques \cite{HarteReview, HFTPaper1, HarteQuad}. Regardless, results of this type are useful because, if $\hat{F}_{ab}$ varies slowly throughout a body's interior, $\mathcal{F}_\tau[ \hat{F}_{ab}, J_c ]$ can be evaluated directly in terms of the multipole moments of $J_a$. Laws of motion for the renormalized momenta then arise which are structurally identical to those satisfied (instantaneously) by a \textit{test body} moving in the effective field $\hat{F}_{ab}$. All $d=4$ self-force results---whether in electromagnetism, scalar field theory, or general relativity---are organized today via similar statements, commonly referred to as Detweiler-Whiting principles \cite{DetweilerWhiting2003, PoissonLR, HarteReview}. We extend such principles here for all $d \geq 3$. If $d$ is even, there is no significant difference with prior work. If $d$ is odd, a crucial difference appears in the form of $\hat{F}_{ab}$.

Before describing this, we first note that a particularly simple Detweiler-Whiting principle holds even in Newtonian gravity \cite{HarteScalar, HarteReview}, where it provides the foundation for Newtonian celestial mechanics: Letting $\phi_S$ denote the self-field associated with a body of mass density $\rho$, each force or torque component $\mathcal{F}_\tau^\mathrm{N} [\phi,\rho]$ may be shown to equal $\mathcal{F}_\tau^\mathrm{N} [\phi - \phi_S,\rho]$; the analog of Eq. \eqref{fVary} holds with $\delta \mathcal{T}^\mathrm{N}_\tau[\rho] = 0$. Self-gravitating bodies thus follow laws of motion which instantaneously coincide with those of a test body in the effective external potential $\hat{\phi} \equiv \phi - \phi_S$. 

We generalize this to electrodynamics by introducing a particular two-point distribution, or ``propagator'' $G_{aa'}(x,x')$ for the vector potential, which plays a role analogous to the Green function $G_\mathrm{N}(\bm{x},\bm{x}') = -|\bm{x} - \bm{x}'|^{-1}$ which appears in the definition for the Newtonian self-field. In particular, our electromagnetic propagator is used to produce the ``effective electromagnetic field''
\begin{equation}
        \hat{F}_{ab}(x) \equiv F_{ab} (x) - 2 \int  \nabla_{[a} G_{a]a'}(x,x') J^{a'}(x') dV'.
        \label{Fhat}
\end{equation}
As it stands, this is merely a definition; it becomes interesting only for propagators with particular properties.

If $d=4$, an appropriate $G_{aa'}$ is known to be singled out by three constraints known as the Detweiler-Whiting axioms \cite{HarteEM, HarteReview, PoissonLR}. As noted also in \cite{HFTPaper1}, these constraints require no modification for any even $d \geq 4$. Eq. \eqref{fVary} may be shown to hold and the Detweiler-Whiting principle remains valid. Explicitly, the propagator implied by this construction is a Green function for Maxwell's equations with the form
\begin{equation}
	G_{aa'} = \frac{1}{2} \left[ U_{aa'} \delta^{(d/2-1)}( \sigma ) +V_{aa'}\,\Theta(\sigma) \right],
	\label{evenD}	
\end{equation}
where $\sigma = \sigma(x,x')$ denotes Synge's world function, one half of the squared geodesic distance between its arguments, and $U_{aa'}$ and $V_{aa'}$ are smooth bitensors, extensions of which also appear in the retarded Green function. The leading-order force acting on a sufficiently small body with charge $q$ and $d$-velocity $\dot{\gamma}^a$ is the usual Lorentz expression $q \hat{F}_{ab}\dot{\gamma}^b$, where $\hat{F}_{ab}$ is obtained from $F_{ab}$ by substituting \eqref{evenD} into \eqref{Fhat}. All dipole and higher-order terms also appear in their usual forms. 

The remainder of this Letter focuses on odd $d$, cases which have not previously been understood (except in the static context \cite{HFTPaper1}). The crux of the complication is that there is no odd-dimensional distribution which satisfies the Detweiler-Whiting axioms. Nevertheless, the formalism used to establish the suitability of those axioms in even dimensions makes it clear that there is considerable freedom to change them. Suppose in particular that a propagator may be found with the properties
\begin{enumerate}
	\item $G_{aa'} (x,x') = 0$ for all timelike-separated $x$, $x'$,

	\item $G_{aa'} (x,x') = G_{a'a} (x',x)$,
	
	\item $G_{aa'}(x,x')$ is constructed only from the geometry and depends quasilocally on the metric,
	
	\item The source $\omega_{d-1}^{-1} \nabla^b \hat{F}_{ab}$ for the effective field is smooth for any monopole point charge moving on a smooth timelike worldline, at least in a neighborhood of that worldline.
\end{enumerate}
The first two of these axioms are shared by the usual Detweiler-Whiting approach. Axiom 1 ensures that the effective momenta are physical in the sense that they depend only quasilocally on the body's state, while Axiom 2 describes a type of reciprocity in the self-field definition associated with $G_{aa'}$ \cite{HarteReview}. The third axiom is similar to one proposed in \cite{HFTPaper1}, and demands more precisely that for any $\psi^a$, the Lie derivative $\mathcal{L}_\psi G_{aa'}(x,x')$ can be written as a functional which depends only on the Lie derivative of the metric, and only in a compact region determined by $x$ and $x'$. If considerations are restricted to a single flat spacetime, Axiom 3 may be simplified by demanding only that $G_{aa'}$ be Poincar\'{e}-invariant. Regardless, Axioms 2 and 3 imply \eqref{fVary}.

Axiom 4 is essentially a point-particle limit of the more physical requirement that simple laws of motion can result only if $\hat{F}_{ab}$ varies slowly throughout a body's interior, at least for sufficiently well-behaved initial data (as is required even for the familiar multipole expansions of Newtonian celestial mechanics). More directly, Axiom 4 ensures that $\mathcal{F}_\tau[\hat{F}_{ab},J_c]$ is generally simpler to evaluate than its bare counterpart $\mathcal{F}_\tau[F_{ab},J_c]$. Although it is satisfied if $G_{aa'}$ is a parametrix for Maxwell's equations, the converse of this statement is false---a distinction which is crucial to our development.

If any $G_{aa'}$ can be found that satisfies the above four axioms, it may be used to write down momenta which obey simple laws of motion in the associated effective field $\hat{F}_{ab}$. The Detweiler-Whiting propagators \eqref{evenD} are explicit examples for all even $d\geq 4$. Before explaining our solution for odd $d$, note that the retarded Green function in those cases has the form $G_{aa'}^\mathrm{ret} = U_{aa'} (-2\sigma)^{1-d/2} \Theta(-\sigma)$ for some smooth $U_{aa'}$, at least within a convex normal neighborhood. Using $g_{aa'}$ to denote the parallel propagator, \begin{equation}
	U_{aa'} = \alpha_d g_{aa'} , \quad \alpha_d \equiv \frac{(-1)^{\frac{1}{2}(d-3)} \Gamma(d/2-1)}{ \sqrt{\pi} \Gamma(\frac{1}{2}(d-1))  }
	\label{Uflat}
\end{equation}
in odd-dimensional Minkowski spacetimes. Whether in Minkowski spacetime or not, Huygens' principle is  violated; signals travel not only along null cones, but also inside of them. Although this occurs also in curved even-dimensional spacetimes, the odd-dimensional case is different in that $G_{aa'}^\mathrm{ret}$ is unbounded even for timelike-separated events. Indeed, it is not even integrable in general. Some care is thus required to define the retarded Green function as a distribution. The correct result may be described by considering $U_{aa'} (-2\sigma)^\lambda \Theta(-\sigma)$ when $\lambda$ is chosen such that the singularity is integrable and then analytically continuing the associated distribution to $\lambda \rightarrow 1-d/2$ \cite{FriedlanderWave, GelfandShilov}. 

Our odd-dimensional propagators are also defined using analytic continuation:
\begin{equation}
        G_{aa'} = \beta_d \lim_{\lambda \rightarrow 1-d/2}  \ell^{2\lambda}\frac{\partial}{\partial \lambda} \left[ \ell^{-2\lambda} U_{aa'} (2 \sigma)^\lambda  \Theta (\sigma)\right].
		\label{EqLogProp}
\end{equation}
These actually constitute a 1-parameter family of propagators, parametrized by an arbitrary lengthscale $\ell > 0$ which ensures that quantity being differentiated is dimensionless. Varying $\ell$ results in effective fields which differ by multiples of a vacuum solution, and although these variations generically  change the effective field and also the force, those forces are associated, via the $\delta \mathcal{T}_\tau$ in \eqref{fVary}, with different effective momenta. Physically, $\ell$ parametrizes different ways to describe the same physical system. As explained in more detail in \cite{HFTPaper1}, true observables do not depend on it.

Noting that $U_{aa'}$ is symmetric in its arguments and quasilocally constructed from the geometry, it is immediately evident that $G_{aa'}$ satisfies our first three axioms. The fourth axiom may be verified by direct computation to hold when
\begin{equation}
	\beta_d = \frac{ (-1)^{\frac{1}{2} (d-1)} }{ 2\pi } .
\end{equation}
Adopting this value establishes a Detweiler-Whiting principle which holds non-perturbatively for extended charges in all odd dimensions. Test body laws of motion hold to all multipole orders, and involve the field $\hat{F}_{ab}$ which is found by substituting \eqref{EqLogProp} into \eqref{Fhat}.

Our propagator $G_{aa'}$ is not a Green function or even a parametrix. Some intuition for it may nevertheless be gained by noting that the derivative with respect to $\lambda$ which appears in its definition evinces a procedure which ``infinitesimally varies $d$.'' This suggests that our map $F_{ab} \mapsto \hat{F}_{ab}$ may reduce to dimensional regularization in a point particle limit, and may provide an underlying physical and mathematical origin for that procedure at least in the present context.

Next, we obtain a point particle limit for the force in the sense described in \cite{GrallaHarteWald, HFTPaper1}. The limiting effective field in this setting is described at leading order by the field of a point mass with charge $q$ moving on a worldline $\gamma(\tau)$ parametrized via its proper time $\tau$. Although there is no conceptual difficulty with considering more general cases, we focus for concreteness on Minkowski spacetime and adopt the inertial coordinates $x^\mu$. Then $g_{\mu\mu'} = \mathrm{diag}(-1,1,\cdots,1)$, and if $F_{ab}$ is assumed to equal the particle's retarded field, a rather involved calculation using \eqref{Fhat},  \eqref{EqLogProp}, and \eqref{Uflat} shows that
\begin{widetext}
\begin{align}
	\hat{F}_{\mu\nu} (\gamma(\tau)) = 2 \alpha_d q \lim_{\epsilon \rightarrow 0^+} \Bigg[ (d-2) \int^{\tau-\epsilon}_{-\infty} \frac{ X_{[\mu}(\tau,\tau') \dot{\gamma}_{\nu]} (\tau')}{ [-X^2(\tau,\tau')]^{d/2} } d\tau' - \sum_{n=0}^{d-4} \frac{ (-1)^n }{ n!  } \bigg( \frac{ \nabla_{[\mu} W_{\nu]}^{(n)} }{ d-3-n } + \frac{1}{\epsilon } \dot{\gamma}_{[\mu} W_{\nu]}^{(n)} \bigg) \frac{1}{\epsilon^{d-3-n}}
	\nonumber
	\\
	~ + \frac{1}{(d-3)!} \left( -\frac{1}{\epsilon} \dot{\gamma}_{[\mu} W_{\nu]}^{(d-3)}  + \nabla_{[\mu} W_{\nu]}^{(d-3)} \ln (\epsilon/\ell) + \frac{1}{2} \partial_\lambda \nabla_{[\mu} W_{\nu]}^{(d-3)} + \frac{1}{(d-2)} \dot{\gamma}_{[\mu} W_{\nu]}^{(d-2)} \right) \Bigg]
	\label{FhatFin}
\end{align}
\end{widetext}
for any odd $d \geq 3$, where $X_\mu(\tau,\tau') \equiv \gamma_\mu(\tau) - \gamma_\mu(\tau')$. If the physical field is not exactly the retarded one, any differences $F_{ab} - F_{ab}^\mathrm{ret}$ are simply added to the right-hand side of \eqref{FhatFin}. All counterterms here---which are derived from \eqref{EqLogProp}, not put in by hand---are expressed in terms of
\begin{align}
	 W_{\mu}^{(n)}(x;\lambda) \equiv \left. \frac{\partial^n}{\partial \tau^n} [ \dot{\gamma}_\mu (\tau) \Sigma^\lambda (x,\tau)] \right|_{\tau=\frac{1}{2} (\tau_{+}+\tau_-)}
	 \label{Wdef}
	\end{align}
and its gradients evaluated at $(\gamma(\tau);1-d/2)$, where $\tau_\pm(x)$ denote the advanced and retarded proper times associated with $x$. The smooth function $\Sigma(x,\tau)$ arises in the factorization $2\sigma(\gamma(\tau), x) = [\tau_+(x) - \tau][ \tau - \tau_-(x)] \Sigma(x,\tau)$, and reduces to unity for an unaccelerated particle. The overall limit here is well-defined, so the effective field is finite on the particle's worldline. More than this, it may be shown that $\hat{F}_{\mu\nu}(x)$ is smooth, as claimed, at least for $x$ in a neighborhood of the particle's worldline; no infinities ever arise. Also note that even though $G_{aa'}$ is not a Green function and the effective field here is not in general a solution to the source-free Maxwell equations, it is source-free for inertially-moving particles. Indeed, it vanishes in these cases.

The leading-order self-force acting on a small charge is now contained in $f_a = q \hat{F}_{ab} \dot{\gamma}^b$. Similarly, the lowest order self-torque acting on a particle with dipole moment $q^{ab} = q^{[ab]}$ follows from $n_{ab} = 2 q^{c}{}_{[a} \hat{F}_{b]c}$. Full laws of motion nevertheless require a centroid or spin supplementary condition, from which a (possibly nontrivial) momentum-velocity relation must be derived. Forces and torques involving higher multipole moments may also be more significant than those which involve the self-force or self-torque. These issues are discussed in detail in \cite{HarteEM, HarteReview, HFTPaper1}. Here, we focus only on the Lorentz force associated with $\hat{F}_{ab}$ in odd dimensions.

Evaluating \eqref{Wdef} and substituting the results into \eqref{FhatFin}, the leading-order $d=3$ self-force is explicitly
\begin{align}
	\label{eq:Lorentz3D}
	f_\mu = 2 q^2 \left( \int^{\tau-\epsilon}_{-\infty} \frac{  X_{[\mu} \dot{\gamma}_{\nu]}' \dot{\gamma}^\nu }{ (-X^2)^{3/2} } d\tau' - \frac{1}{4} \ln(\epsilon/e \ell) \ddot{\gamma}_\mu \right),
\end{align}
where the limit $\epsilon \rightarrow 0^+$ has been left implicit and $e$ is the base of the natural logarithm. Different choices for $\ell$ result in different momenta, and thus different forces. It is clear in this context that the corresponding momentum differences involve only changes in magnitude, i.e., differing effective masses. The equivalent expression for $d=5$ is somewhat more complicated: Letting $h_{\mu\nu} \equiv \eta_{\mu\nu} + \dot{\gamma}_\mu\dot{\gamma}_\nu$ denote the spatial projection operator,
\begin{align}
	f_\mu = - q^{2} \Bigg( \int_{-\infty}^{\tau-\epsilon} \frac{  3 X_{[\mu} \dot{\gamma}_{\nu]}' \dot{\gamma}^\nu }{ (-X^2)^{3/2} } d\tau'+h_{\mu\nu} \bigg[ \frac{3\ddot{\gamma}^{\nu} }{8\epsilon^{2}} - \frac{\dddot{\gamma}^\nu}{2\epsilon} \nonumber\\
	~ - \frac{3}{16} (\ddddot{\gamma}^\nu - \frac{3}{2} |\ddot{\gamma}|^2 \ddot{\gamma}^\nu ) \ln ( \epsilon/e^{1/3} \ell) - \frac{|\ddot{\gamma}|^2 \ddot{\gamma}^\nu  }{32} \bigg] \bigg).
	\label{eq:Lorentz5D}
\end{align}
Changing $\ell$ thus renormalizes more than just the mass; it affects both the direction and magnitude of the 5-momentum.

A force similar to \eqref{eq:Lorentz5D} has recently been obtained using the methods of effective field theory \cite{Chad}, and in that context, $\ell$ appears as a free parameter in a dimensional regularization procedure. It is interpreted there as a lengthscale over which measurements may be performed, an interpretation which does not arise in our framework.

We now consider the nonrelativistic limits of the $d=3$ and $d=5$ self-forces. The leading-order terms are purely spatial, and in three dimensions, repeatedly integrating by parts results in 
\begin{equation}
	\bm{f}(\tau) = -\frac{1}{2} q^2 \int_{-\infty}^\tau \dddot{\bm{\gamma}}(\tau')  \ln \left ( \frac{ \tau-\tau'}{ e^{1/2} \ell} \right)  d\tau' ,
	\label{fNonrel}
\end{equation}
where it has been assumed that the particle's motion is inertial in the distant past. The $d=5$ case is similar, but with the third derivative of the particle's position replaced by the fifth derivative (and with different constants). 

Focusing on three dimensions for definiteness, the self-force is seen to depend on the past history of the particle's jerk, with a weighting factor for this history which \textit{increases} in the increasingly distant past. If a charge is initially stationary, is subjected to a brief impulse near $\tau=\tau_0$, and moves inertially thereafter, $\bm{f}(\tau) = -\frac{1}{2} q^2 \Delta \bm{v}/(\tau-\tau_0)$ at late times, where $\Delta \bm{v}$ denotes the velocity change associated with the impulse. This scenario requires a small external force to be applied after the initial impulse in order to counteract the self-force. If that force is not applied, a freely-evolving particle would slow down after the initial impulse. Using the above $\bm{f}(\tau)$ to estimate that slowdown---which is analogous to the reduction-of-order approximations commonly applied in four-dimensions---the velocity change would appear to grow logarithmically at late times. Although this unphysical conclusion signals that it is inappropriate to approximate the force by the above expression, it does illustrate the remarkably strong memory of the $d=3$ self-force. Attempting to alternatively estimate the slowdown by cutting off the force integral before the external force is applied results in charges always returning asymptotically to rest. This approximation is also not well-controlled, although it does suggest that the self-field sourced in the distant past dissipates so slowly as to provide a kind of eternal rest frame. The actual endstate is nevertheless unclear. Faster decay rates in higher dimensions considerably simplify the situation.

Another example which is worth noting is that of harmonic motion. Suppose that an external force is applied such that $\bm{\gamma}(\tau)$ varies sinusoidally. Although the falloff conditions used to derive \eqref{fNonrel} are violated in this case, applying fewer integrations by parts allows a well-defined self-force to be computed. Regardless of frequency, the phase of the $d=3$ self-force is advanced with respect to the phase of the particle's acceleration. Applying this for a charge in a fixed circular orbit with angular velocity $\omega$, the nonrelativistic self-force reduces to
\begin{equation}
	\bm{f}(\tau) = \frac{1}{4}q^2 \left[  \left( 1 + 2 \gamma_E + \ln (\omega\ell)^2 \right) \ddot{\bm{\gamma}}(\tau) - \pi \omega \dot{\bm{\gamma}}(\tau) \right] ,
\end{equation}
where $\gamma_E$ denotes the Euler-Mascheroni constant. While work is performed only by the term proportional to the velocity, the term proportional to acceleration provides an effective mass which depends on $\ln \omega$. The analogous force in five dimensions is essentially the same except for an additional factor of $\omega^2$.

We end by comparing the results presented here to others which have been suggested in the literature. Unlike our approach in which all results follow from first principles and in which no infinities arise, other proposals may be summarized as heuristic attempts to directly regularize point-particle self-fields \cite{Galtsov, Galakhov, Kazinski} or expressions for the momentum associated with such fields \cite{Yaremko}. In at least one case, the claimed force law is IR-divergent; see Eq. (4.4) in \cite{Galtsov}. Other claimed force laws involve counterterms which depend on the particle's entire past history \cite{Kazinski, Yaremko}, laws which could arise only if a particle's momenta depended on its state in a highly nonlocal manner---a physically-unacceptable option. Another result predicts a time-varying mass even at lowest order \cite{Yaremko}. These conclusions are qualitatively different from ours.

\bibliographystyle{apsrev4-1}
\bibliography{selfforce}

\end{document}